\documentclass[useAMS,usenatbib]{mn2e}
\usepackage[dvips]{graphicx,color}
\usepackage[latin1]{inputenc}
\usepackage{graphics}
\usepackage{amsfonts}
\usepackage{amsmath}
\usepackage{multicol}
\usepackage{layout}
\usepackage{amssymb}
\usepackage{epsfig}
\usepackage[a4paper]{hyperref}  
\DeclareGraphicsExtensions{.eps}  
\title[The Population of Dark Matter Subhaloes]
{The Population of Dark Matter Subhaloes: Mass Functions
and Average Mass Loss Rates}
\author[C. Giocoli, G. Tormen \& F. C. van den Bosch]{Carlo Giocoli$^{1}$, 
Giuseppe Tormen$^{1}$ \& Frank C. van den Bosch$^{2}$
\thanks{E-mail: 
\href{mailto:carlo.giocoli@unipd.it}{carlo.giocoli@unipd.it},
\href{mailto:giuseppe.tormen@unipd.it}{giuseppe.tormen@unipd.it}
\href{mailto:vdbosch@mpia-hd.mpg.de}{vdbosch@mpia-hd.mpg.de}.}\\  
\\
$^{1}$Dipartimento di Astronomia, Università degli Studi di Padova, 
Vicolo dell'osservatorio 2 I-35122 Padova, Italy \\
$^{2}$ Max-Planck-Institute for Astronomy,  Konigstuhl 17, 69117 
Heidelberg, Germany \\
}
\begin{document}
\date{}
\pagerange{\pageref{firstpage}--\pageref{lastpage}} \pubyear{2007}
\maketitle
\label{firstpage}

\begin{abstract}
Using a cosmological N-Body simulation and a sample of re-simulated
cluster-like haloes, we study the mass loss rates of dark matter
subhaloes, and interpret the mass function of subhaloes at redshift
zero in terms of the evolution of the mass function of systems
accreted by the main halo progenitor (hereafter called the `unevolved
subhalo mass function').  When expressed in terms of the ratio between
the mass of the subhalo at the time of accretion, $m_v$, and the
present day host mass, $M_0$, the unevolved subhalo mass function is
found to be universal, in that it is independent of the mass of the
host halo. However, the subhalo mass function at redshift zero
(hereafter called the `evolved subhalo mass function') clearly depends
on $M_0$, in that more massive host haloes host more subhaloes.  In
order to relate the unevolved and evolved subhalo mass functions, we
measure the subhalo mass loss rate as a function of host mass and
redshift.  We find that the average, specific mass loss rate of dark
matter subhaloes depends mainly on redshift, with only a very weak
dependence on the instantaneous ratio between the mass of the subhalo,
$m_{sb}$, and that of the host halo at that time $M_v$. In fact, to
good approximation, subhalo masses `decay' exponentially, with a
decay-time that is proportional to the instantaneous dynamical time of
the host halo.  Combined with the fact that more massive haloes
assemble later, these results suggest a pleasingly simple picture for
the evolution and mass dependence of the evolved subhalo mass
function.  Less massive host haloes accrete their subhaloes earlier,
which are thus subjected to mass loss for a longer time.  In addition,
their subhaloes are typically accreted by denser hosts, which causes
an additional boost of the mass loss rate.  To test the
self-consistency of this picture, we use semi-analytical merger trees
constructed using the extended Press-Schechter formalism, and evolve
the subhalo populations using the average mass loss rates obtained
from our simulations.  The resulting subhalo mass functions are found
to be in good agreement with the simulations.  Our model can be
applied to semi-analytical methods of galaxy formation, to accurately
follow the time evolution of subhalo masses.
\end{abstract}


\section{Introduction}

Understanding structure  formation  is a  fundamental  topic in modern
cosmology.  In  the     current  $\mathrm{\Lambda  CDM}$   concordance
cosmology, the   matter density of the Universe   is dominated by cold
dark matter (CDM),  whose  gravitational  evolution  gives rise to   a
population  of virialized dark matter  haloes spanning a wide range of
masses.    Numerical  simulations of    structure formation  in  a CDM
universe predict that these dark matter haloes contain a population of
subhaloes, which  are the remnants of  halos accreted by the host, and
which are eroded by the  combined effects of gravitational heating and
tidal stripping in the potential well of the main halo.

Understanding the evolution of the  subhalo mass function, as function
of   cosmology,  redshift, and  host   halo   mass, is  of   paramount
importance,   with  numerous applications.     For one,  subhaloes are
believed  to host  satellite galaxies,   which  can thus   be used  as
luminous tracers of the subhalo population. In particular, linking the
observed abundances of satellite galaxies to the expected abundance of
subhaloes,  provides  useful  insights  into   the physics of   galaxy
formation     \citep[e.g.,][]{mooreetal99,bull00,som02,ketal04b,vo06}.
Studies along   these lines  indicate that  galaxy   formation becomes
extremely inefficient in  low mass haloes,  and suggest that there may
well be  a large population  of low  mass  subhaloes with  no  optical
counterpart     \citep[e.g.,][]{mooreetal99,setal03,ketal04a}.      In
principle, though,  these truly `dark'  subhaloes  may potentially be
detected via $\gamma$-ray emission due to  dark matter annihilation in
their                              central                       cores
\citep{setal03ann,bertone06,kou06,pieri07,diemand07}, or  via    their
impact   on the   flux-ratio   statistics of  multiply-lensed  quasars
\citep[e.g.,][]{mm01,DalKoc02}.   Alternatively, these techniques  may
be used to constrain  the  abundance of  subhaloes, which in  turn has
implications for  cosmological parameters  and/or the nature   of dark
matter.   The evolution   of the subhalo   mass  function is also   of
importance   for  the   survival  probability    of   disk    galaxies
\citep[e.g.,][]{TO92,betal04,Stewetal07} and even has implications for
direct detection  experiments of dark  matter  \citep[e.g.,][]{Goer07}
Finally, understanding the rate at  which dark matter subhaloes  loose
mass has  important implications  for their dynamical  friction times,
and       thus    for     the      merger      rates  of      galaxies
\citep[e.g.,][]{betal02,ZB03,TB04}.

Despite significant  progress  in the   last years, there  are   still
numerous issues that are insufficiently understood.   What is the mass
function of haloes accreted onto the main progenitor  of a present day
host halo?  How do  the orbits and masses of  subhaloes evolve as they
are subjected to dynamical friction, tidal forces and close encounters
with other  subhaloes?  How does this  depend on the properties of the
host  halo?   In   this  paper we    address   these  questions  using
high-resolution numerical simulations.  We trace back the evolution of
self-bound  substructures identified in present-day  host haloes up to
the point where they are first accreted by the  main progenitor of the
host  halo.  Using  this method we   are able to link the  present-day
population of subhaloes to the merging history of the host system.  We
will  show that  larger systems,   forming  at lower redshifts  and so
accreting  their satellites  more  recently, contain  at the end  more
subhaloes than smaller hosts \citep[see also][]{getal04,vtg05}.

In  Section~\ref{simulation}  we describe  the  simulations used.   In
Section~\ref{postprocessing}  we present  the  algorithms employed  to
identify the  haloes and  to follow their  merging history  trees.  In
Section~\ref{massaccreted}  we  show how  the  unevolved subhalo  mass
function is constructed from the  merger tree of present-day halos and
suggest an analytical fit.   In Section~\ref{subhaloes} we explain how
subhaloes   are  identified   at   the  present   time.   In   Section
\ref{massloss}  we calculate  the mass  loss rate  for  subhaloes, and
characterize its  dependence on  host halo mass  and on  redshift.  In
Section~\ref{mcsim} we present  Monte Carlo simulations that reproduce
the  subhalo  mass function  measured  in  the  $N$-body simulations.  
Finally, in Section~\ref{conclusion} we summarize our results and draw
some conclusions.

\section{The Simulations}
\label{simulation} 

To study the evolution of the subhalo population of dark matter haloes
we use two different types of  simulations: a set of 48 massive haloes
that  have been  extracted from  a large  cosmological  simulation and
resimulated  at  much higher  resolution,  and  a  set of  two  large,
cosmological  simulations that probe  a much  larger dynamic  range in
host halo masses.

\subsection{Resimulations}

Our sample  of $48$ resimulated  dark matter haloes is  extracted from
ten  high-resolution   $N$-body  resimulations  of   galaxy  clusters,
containing  $512^3$ particles  in  a cube  $479\,\mathrm{Mpc}/h$ on  a
side. All simulations assume  a flat $\mathrm{\Lambda CDM}$ model with
$\Omega_{0}=0.3$,  $h=0.7$,   $\sigma_{8}=0.9$  and  $\Omega_{b}=0.04$
\citep{ysd01}.  The masses of  the haloes  selected to  be resimulated
cover     the      range     $5.1     \times     10^{13}\,-\,2.3\times
10^{15}\,M_{\odot}/h$ at redshift $z=0$.

\begin{figure*}
 \begin{center}
   \includegraphics[width=\hsize]{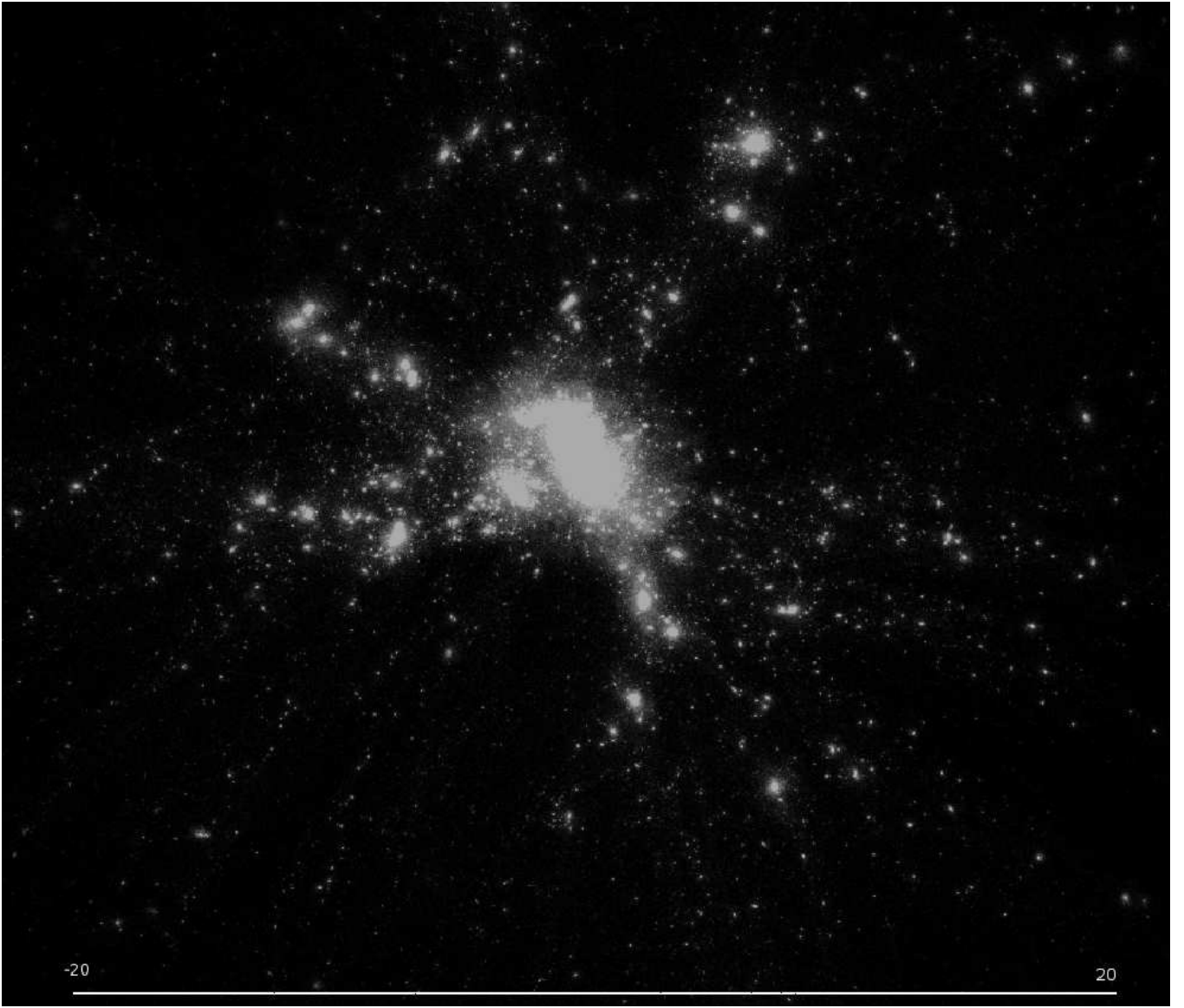}
   \caption{A cluster of the resimulated sample resolved with $2
     \times 10^{6}$ particles within the virial radius and $6 \times
     10^{6}$ high resolution particles within
     $40\,\mathrm{Mpc}/h$.\label{g8}}
 \end{center}
\end{figure*}

For  the  resimulations,  we  adopt  a  particle  mass  of  $1.3\times
10^{9}\,M_{\odot}/h$ and a gravitational softening length of $\epsilon
=  5$ kpc/$h$ (Plummer  equivalent).  The  initial conditions  for the
resimulation are generated with higher mass and force resolution using
the  Zoomed Initial  Condition  technique \citep[ZIC,][]{tbw97}:  halo
Lagrangian regions are populated with  a larger number of less massive
particles, and  additional small-scale  power is added  appropriately. 
The new initial conditions are evolved using the Tree-SPH code Gadget2
\citep{springg2} from redshift $z=60$  to the present time, using dark
matter particles only.  We study these resimulations using $88$ output
times equally spaced between  $z=10$ and $z=0$.  Figure~\ref{g8} shows
an example of one  of these resimulated cluster-sized haloes, embedded
in  its surrounding  large-scale structure.   The cluster  is resolved
with more  than one  million particles within  its virial  radius, and
there  are  roughly  $6  \times  10^6$  high  resolution  dark  matter
particles inside $20$ Mpc/$h$. See \citet{dvbt} for further details.

\subsection{Cosmological N-Body Simulations}

We will also  make use of two cosmological  $N$-body simulations.  The
first is the so called "`GIF2"' simulation \citep{getal04}, a periodic
cube   of   side  110   $\mathrm{Mpc}/h$   assuming  the   concordance
$\mathrm{\Lambda}$CDM model  ($\Omega_{\Lambda}$, $\Omega_b h^2$, $h$,
$\sigma_8$)=(0.7, 0.0196,  0.7, 0.9). The  index of the  initial power
spectrum as be  chosen $n=1$, with the transfer  function produced by
CMBFAST \citep{sz96}.
GIF2  contains $400^{3}$  dark matter  particles, each  of  mass $1.73
\times  10^{9}\,M_{\odot}   /h$.   We  will  use   $50$  output  times
logarithmically  spaced between  $z=10$ and  $z=0$, which  suffices to
construct accurate  halo and subhalo  merging history trees  using the
method of  \citet{tmy04}.  The numerical  data of the  GIF2 simulation
are publicly  available at \href{http://www.mpa-garching.mpg.de/Virgo}
{\texttt{http://www.mpa-garching.mpg.de/Virgo}}.

Finally, as a consistency check, part of the analysis done on GIF2 was
repeated on  the lower resolution GIF  box ($\mathrm{\Lambda}$CDM 
each of mass  $1.4 \times 10^{10}\,M_{\odot} /h$), which  has the same
cosmological parameters of the GIF2 simulation. See \citet{kcdw99} for
a detailed description of this simulation.

\begin{table*}
\centering
\begin{tabular}{|l || c || c || c || c || c || c || c |} \hline
Sim. name & $11.5$-$12$ & $12$-$12.5$ & $12.5$-$13$ & $13$-$13.5$ &
$13.5$-$14$ & $14$-$14.5$ & $> 14.5$  \\ \hline \hline 
Resimulations & - & - & - & - & $21$ & $17$ & $10$  \\ \hline 
GIF & - & - & $2693$  & $971$ & $290$ & $99$ & $16$  \\ \hline 
GIF2 & $8305$ & $3349$ & $1186$ & $461$ & $127$ & $35$ & $4$  \\ \hline 
\end{tabular}
\caption{ The  number of haloes in  each logarithmic mass  bin for the
different simulations. For  GIF \& GIF2 we consider  all haloes with a
least more that $200$ particles within their virial radius at redshift
zero and whose  main progenitor never has a  virial mass exceeding the
final value  by more than ten  percent. For the  resimulated haloes we
follow  the merger  tree and  the satellites  populations for  all the
haloes with more than $40000$ particles at the present time.}
\label{tabsum}
\end{table*}


\subsection{Halo-Finder \& Merger History Tree} 
\label{postprocessing}

We  adopt the spherical  overdensity criterion  to identify  haloes at
each  simulation output  time  (also called  "`snapshot"').  For  each
snapshot we estimate the local  dark matter density at the position of
each  particle  by  calculating  the  distance to  the  tenth  closest
neighbor.   We   assign   to    each   particle   a   local   density
$\rho_{i,\mathrm{DM}}   \propto  d_{i,10}^{-3}$,  sort   particles  in
density  and take  as centre  of the  first halo  the position  of the
densest particle.  We then grow a sphere of matter around this centre,
and  stop when  the mean  density within  the sphere  falls  below the
virial value appropriate for  the cosmological model at that redshift;
for  the  definition  of  virial  density  we  adopted  the  model  of
\citet{eetal96}; for example, at  redshift $z=0$ the virial density is
$\rho_v =  324 \rho_b$, with  $\rho_b$ the mean background  density of
the universe.

At this point  we assign all particles within the  sphere to the newly
formed halo, and remove them from  the global list. We take the centre
of the  next halo at  the position of  the densest particle  among the
remaining ones, and grow a  second sphere.  We continue in this manner
until all  particles are screened.   We include in our  catalogue only
haloes  with  at  least  $10$  particles  within  the  virial  radius;
particles not  ending up  in any halo  are considered as  "`field"' or
"`dust"' particles.

We  then  build  the  merging  history  tree for  all  haloes  in  the
simulation  (or  resimulation)  using   the  halo  catalogues  at  all
snapshots, separated by redshift intervals ${\rm d}z_{i}$, as follows.
Starting from  each halo  at $z=0$, we  define its progenitors  at the
previous output,  $z={\rm d}z_{1}$, as all haloes  containing at least
one  particle  that at  $z=0$  will belong  to  that  halo. The  `main
progenitor'  at $z={\rm  d}z_{1}$ is  defined as  the  progenitor that
provided the largest mass contribution  to the halo at $z=0$.  Next we
repeat  the  same procedure,  now  starting  at  $z={\rm d}z_{1}$  and
considering progenitors at $z={\rm d}z_{1}  + {\rm d}z_{2}$, and so on
backward  in time,  always  following the  main  progenitor halo.  The
resulting merger tree consists of  a main trunk, which traces the main
progenitor  back in  time,  and  of `satellites',  which  are all  the
progenitors  which,  at  any   time,  merge  directly  onto  the  main
progenitor.

In the  following analysis we  only consider haloes at  redshift $z=0$
whose main progenitor  at any redshift has a  virial mass $M_v(z)$ not
exceeding  the  final value  $M_v(z=0)\equiv  M_0$  by  more than  ten
percent.  In  fact, any $M_v(z)  / M_0$ significantly larger  than one
corresponds to  an incomplete merger  in which two haloes  first merge
but  subsequently   split  again.    This  typically  occurs   when  a
(relatively) small halo pass through  a larger one. Since these events
complicate our analysis,  and their occurrence is only  low, we decided
to eliminate such merger histories.

For all simulations we split the halo samples at $z=0$ in mass bins of
width   $\mathrm{d}\log(M)=0.5$,   with   a   minimum   mass   roughly
corresponding to $200$ particles within  the virial radius for GIF and
GIF2, and to $40000$ particles for the resimulations.  The actual mass
bins for each run are listed in Table~\ref{tabsum}.

\begin{figure*}
\begin{center}
   {\includegraphics[width=7.9cm]{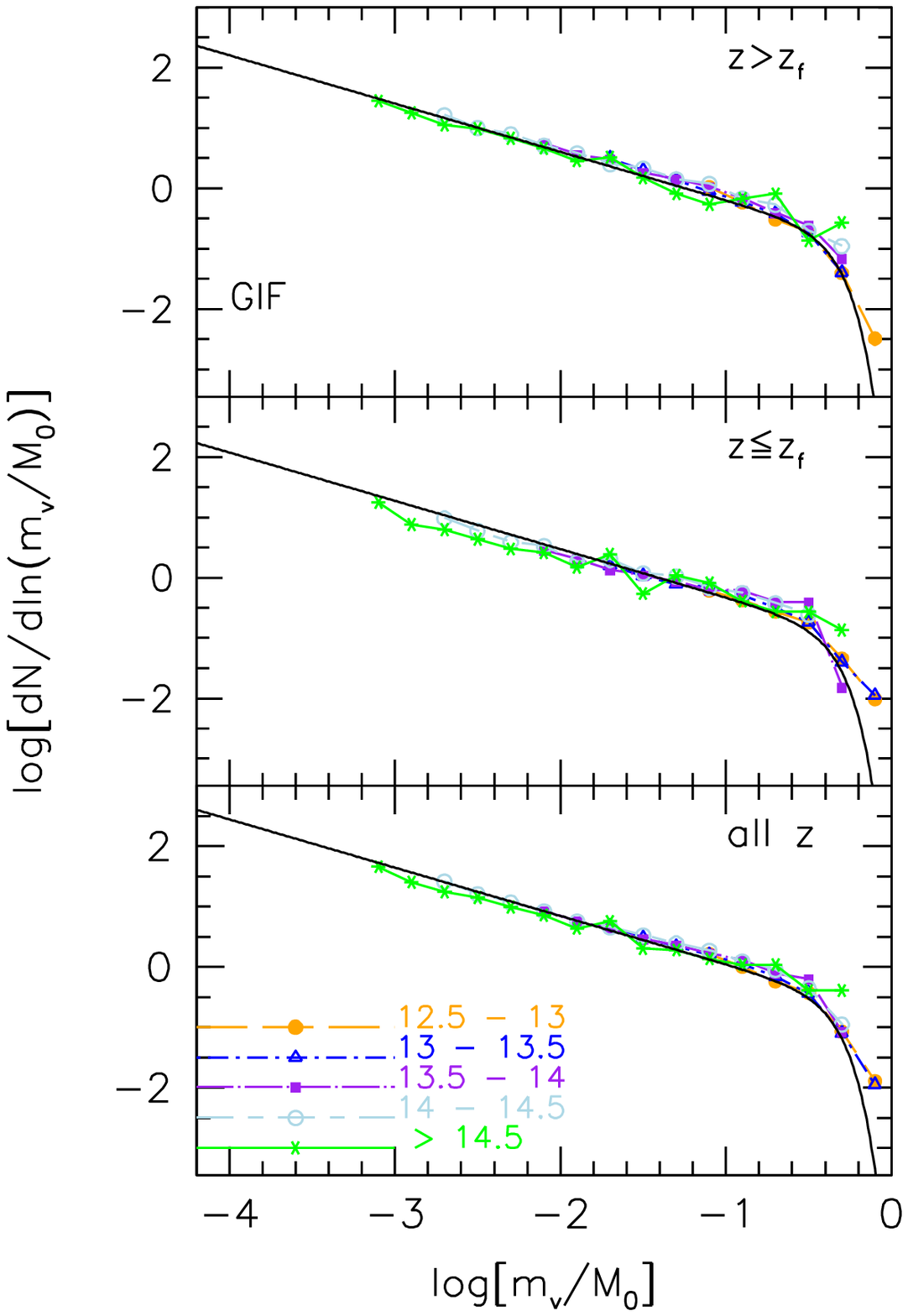}}
   \hspace{5mm}
   {\includegraphics[width=7.9cm]{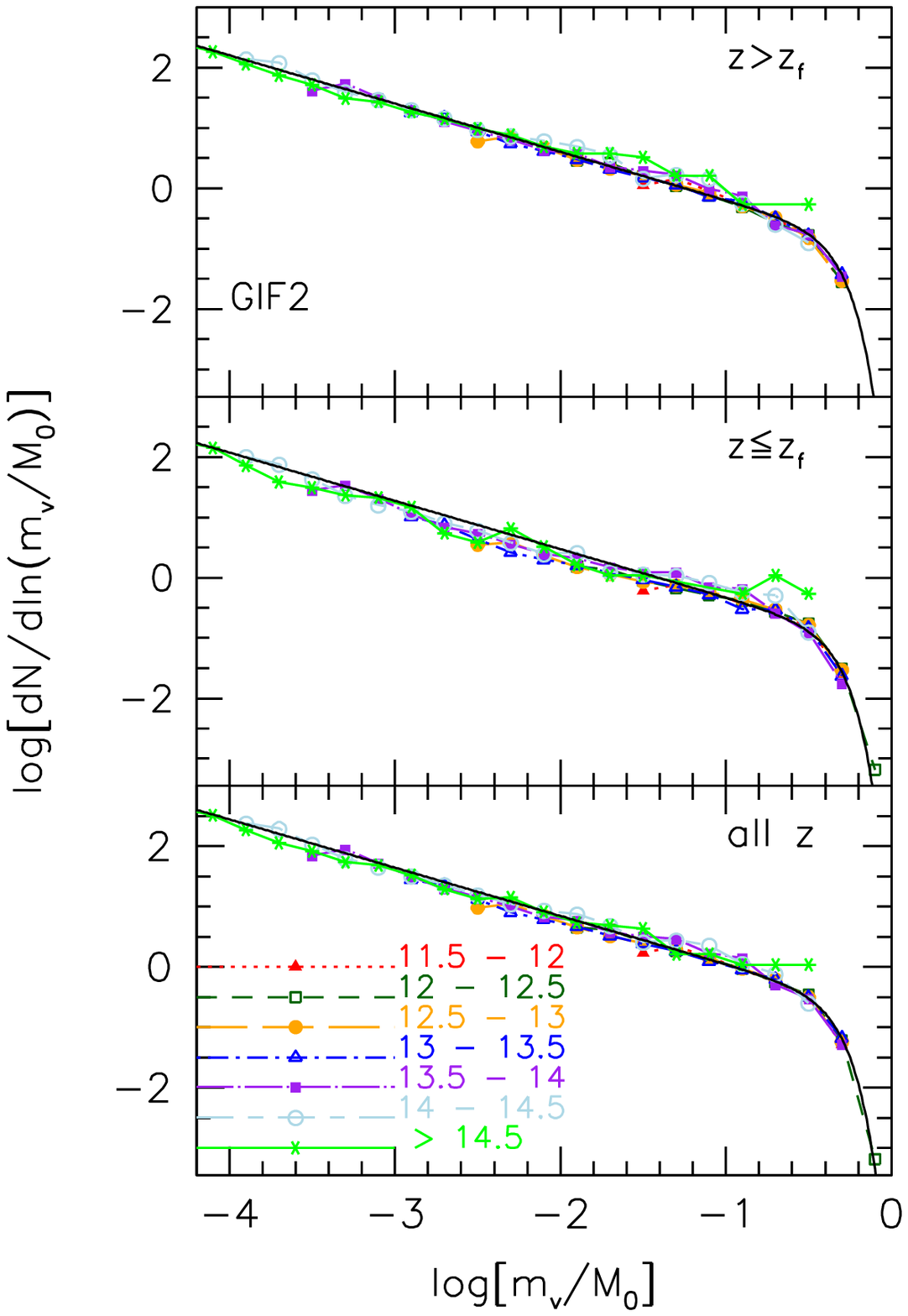}} 
   \vspace{5mm} 
   \caption{Mass functions  of accreted satellites  (unevolved subhalo
     mass functions). In  the panels the various data  points and line
     types refer  to different present-day  host halo masses.   In the
     figures the the bounds of the  mass bins are expressed in unit of
     $\log(M_{\odot}/h)$.   The  solid  lines  represent  the  fitting
     function to the distributions: equation (\ref{unshmfeq}) (see the
     main  text  for  more  details).   Note  that  we  only  consider
     subhaloes  that at  $z=0$  contribute at  least  $50\%$ of  their
     mass. This ensures that at $z=0$ their center of mass lies within
     the  virial radius  of the  host.  \emph{Top}:  Unevolved subhalo
     mass function accreted before the formation redshift $z_f$ of the
     host halo (defined as the  earliest redshift when the mass of the
     main  progenitor exceeds half  the final  mass).  \emph{Center}:
     Same as above, but only counting satellites accreted after $z_f$.
     \emph{Bottom}:  Same  as   above,  but  now  counting  satellites
     accreted at any redshift. \label{unshmf}}
\end{center}
\end{figure*}


\section{Unevolved subhalo mass function}
\label{massaccreted}

Starting from each halo at $z=0$,  we trace its merger history back in
time  and  register all  its  satellites,  i.e.   all haloes  directly
accreted by the halo main progenitor  at any output time.  In order to
remove subhaloes  that at $z=0$ reside  outside the host  due to their
elongated orbits (and so do  not contribute to the subhalo population)
we only consider satellites which donate at least $50\%$ of their mass
to the final halo.  Let  $n(m_v/M_0,z)$ be the number of satellites of
virial mass $m_v$,  accreted at redshift $z$ by a  host halo with mass
$M_0$ at redshift zero.  Integrating this expression over the redshift
interval  $z_1  \leq  z \leq  z_2$,  we  obtain  the total  number  of
satellites of mass $m_v$ accreted by the main progenitor during that
interval,
\begin{equation}
\mathrm{N} \left( \frac{m_v}{M_0}\right) = 
\int_{z_1}^{z_2} n\left( \frac{m_v}{M_0},\zeta \right)
\mathrm{d} \zeta\,
\label{eq:nofz}
\end{equation}
which we  term the \emph{unevolved subhalo mass function}.

In Figure~\ref{unshmf} we plot the unevolved subhalo mass function for
different redshift intervals,  as measured in  the GIF (left) and GIF2
(right) simulations.  The data points refer to  different mass bins of
the  parent haloes at redshift $z=0$,  as indicated.  As stated above,
we only   considered satellites that contributed   at  least $50\%$ of
their mass to the final ($z=0$)  host. Setting $z_1=0$ and $z_2=z_{\rm
  max}$,  which is the maximum  redshift available in the simulations,
we  obtain the  total, unevolved  subhalo mass functions  shown in the
lower panels of Figure~\ref{unshmf}. Note that  there is no indication
for any significant dependence on $M_0$, indicating that the unevolved
subhalo mass function is indeed  universal, as found by \citet{vtg05}.
After  some experimenting, we  find  that the  unevolved subhalo  mass
function is well fitted by
\begin{equation}
\frac{\mathrm{d}\mathrm{N}}{\mathrm{d}\ln(m_{v}/M_{0})}= N_{0}
x^{-\alpha} \mathrm{e}^{-6.283 \,
x^{3}},\;\;\;\,x=\Big{|}\frac{m_{v}}{\alpha M_{0}} \Big{|}
\label{unshmfeq}
\end{equation}
with $\alpha=0.8$ and $N_{0}=0.21$. This fitting function is indicated
by solid black lines in the lower panels.
  
In the upper and middle panels of Figure~\ref{unshmf} we show the mass
functions  of the satellites accreted  at redshifts larger and smaller
than $z_f$,   respectively.    Here $z_f$ is    the so-called assembly
redshift, defined as  the {\it highest}  redshift at which the mass of
the   main progenitor $M_v(z)$ exceeds   half  the final value, (i.e.,
$M_v(z) >  M_0/2$).  Once again, the  results for different  host halo
masses are   indistinguishable,   and  are  extremely  well   fit   by
equation~(\ref{unshmfeq}) with  $\alpha=0.8$.      The  normalization,
$N_0$, however,  needs to be adjusted.  Naively one  would expect that
both  mass functions should  have  a normalization $N_0/2$.   However,
because of the discreteness of the  merger histories, the average mass
at the formation redshift, $M(z_f)$,  is actually slightly larger than
$M_0/2$.   \citet{st04}  have shown that,  for  the spherical collapse
case and assuming a white noise power  spectrum, the mass at formation
has a  distribution  (eq.[4] of    their paper)  with   a mean   value
$\mu_{ST04}M_0 = (0.586 \pm 0.005)  M_0$.  Here, combining haloes from
GIF and GIF2, we find a mean formation mass $\bar{\mu}_{GIF+GIF2}M_0 =
(0.572\pm 0.001) M_0$.   The fact that our  results are somewhat lower
owes  to the fact that the  distribution of $\mu$  depends (weakly) on
the power  spectrum  \citep[see][]{st04}.  The  normalizations of  the
\emph{unevolved subhalo mass  function} accreted before  $N_{0,b}$ and
after $N_{0,a}$ the formation redshift are therefore:
\begin{eqnarray}
N_{0,b} &=& \bar{\mu} N_0 = 0.572 N_0\,,\\ 
N_{0,a} &=& (1- \bar{\mu}) N_0 = 0.428 N_0\,.
\end{eqnarray}
These  are the  normalizations that  we have  adopted for  the fitting
functions shown in the upper and middle panels of Figure~\ref{unshmf}.


\section{Evolved subhalo mass function}
\label{subhaloes}

The evolved  subhalo mass function at  any redshift $z$  is built from
all  the satellite  haloes  accreted  by the  main  progenitor at  all
redshifts larger than $z$, where for each satellite we compute at each
redshift its  self-bound mass $m_{sb}(z)$.   Operationally, we perform
the following steps:
\begin{itemize}
  
\item given a satellite halo, we identify its merging redshift, $z_m$,
  defined  as the  latest redshift  when  the satellite  was still  an
  isolated halo, just before it was accreted by the main progenitor;
\item we calculate the position of its center using the "moving center
  method"' \citep{tbw97}, i.e.  by  repeated calculation of its center
  of  mass using  smaller and  smaller radii  to identify  the subhalo
  densest core;
\item we compute the subhalo tidal radius - as in \citet{tetal98};
\item  we evaluate  the binding  energy  of each  subhalo particle  by
  summing its potential energy  (calculated using all particles inside
  the  tidal  radius)  and  its  kinetic energy  (using  its  residual
  velocity with respect to the average value inside the tidal radius);
\item  we  remove all  particles  with  positive  binding energy,  and
  iterate  the  previous  steps  until  the  self-bound  subhalo  mass
  converges.
\end{itemize} 

With these data in hand, we  can follow the time evolution of the self
bound mass  of each  subhalo, snapshot by  snapshot, from  the merging
redshift $z_{m}$ to the present  time $z=0$.  In the following Section
we will use this information  - gathered from the resimulated haloes -
to estimate the subhalo mass-loss rates at all redshifts.

\begin{figure}
  \centering \includegraphics[width=\hsize]{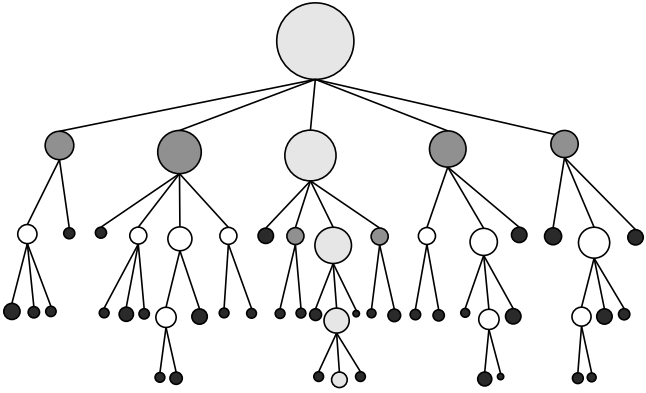}
  \caption{Schematic representation of the merging-history-tree of an
    halo.  Solid  light gray  circles  connected  on  the parent  halo
    represent the  main branch  of the tree.  Solid dark  gray circles
    indicate   satellites.   Solid   black  circles   indicate  leaves
    progenitors. See the main text for explanation.
  \label{mergertreerap}}
\end{figure}

In  Figure~\ref{mergertreerap} we show  a schematic  representation of
the merging history tree of a  halo.  Time runs upward, with the final
halo  depicted at  the  top.   Light gray  circles  indicate the  main
progenitor at each time, whose history defines the `main trunk' of the
tree. Dark  gray circles  indicate satellite haloes,  i.e.  progenitor
haloes  accreted  directly  by  the  main  trunk  of  the  tree.   The
progenitors  indicated  by  black  circles  are the  `leaves'  of  the
merging-history tree,  and reflect those progenitors whose  mass is of
the order  of the resolution  of the tree;  i.e., for these  haloes no
progenitor can  be identified  in the simulation  at earlier  outputs. 
Satellites  are either  `leaves', or  they have  progenitor  haloes at
earlier  outputs, which  in principle  give  rise to  a population  of
sub-subhaloes,  etc.    In  this  paper   we  do  not   consider  this
substructure of subhaloes, though we intend to address their evolution
in a forthcoming paper (Giocoli et al. in preparation).

\begin{figure*}
\begin{center}
   \includegraphics[width=\hsize]{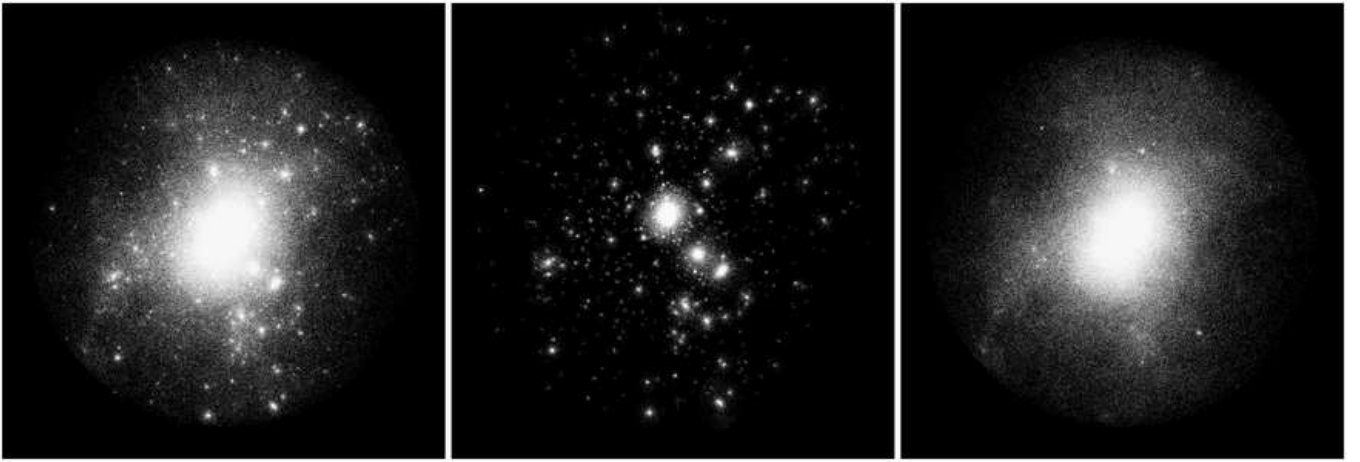}
   \caption{\label{psub}Subhalo population. Left: all particles
     composing  the most  massive  halo  found at  $z=0$  in the  GIF2
     simulation; the  virial mass for this  halo is $M_v  = 1.8 \times
     10^{15}\,M_{\odot}/h$,   resolved  by   more  than   one  million
     particles.   Center:  particles bound  to  subhaloes at  redshift
     $z=0$.  Right:  particles bound  to  the  main  halo but  not  to
     subhaloes.}
\end{center}
\end{figure*}

As an  example, Figure~\ref{psub} shows the subhalo  population of the
most massive  halo found  at $z=0$ in  the GIF2 cosmological  run. The
left panel shows all particles inside the halo virial radius $R_v$. In
the  middle panel only  those particles  are shown  that are  bound to
subhaloes located within $R_v$, while  the right panel shows all other
`field' particles,  which are bound to  the main halo, but  not to any
subhalo.

\begin{figure}
\begin{center}
   \includegraphics[width=7.9cm]{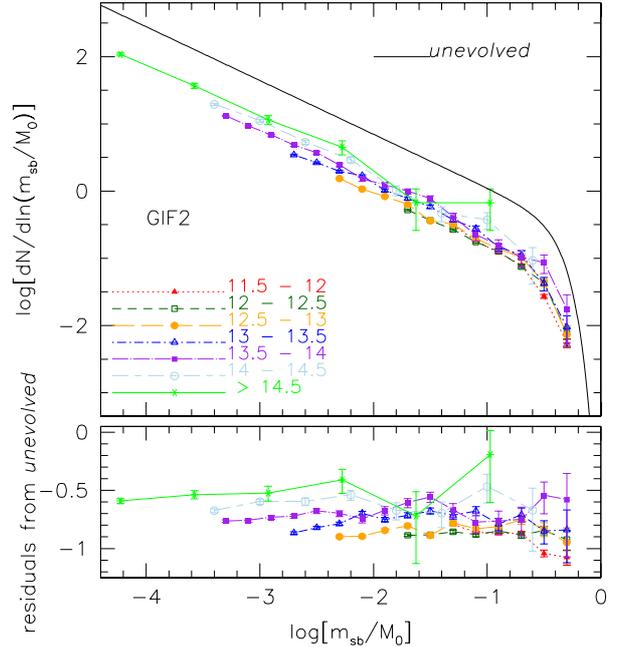}
   \caption{\label{subhmf1}Subhaloes  mass function of  the self-bound
     particles  of the  haloes  accreted  by the  main  branch of  the
     merger-history-tree of an halo,for  GIF2 simulation.  In the plot
     it has  been considered all  satellites with a distance  from the
     center of  the host  halo less then  the virial radius.   We also
     plot  the unevolved distribution:  equation~(\ref{unshmfeq}). The
     different data points and line  types used are the same of Figure
     \ref{unshmf}.}
\end{center}
\end{figure}

In  Figure~\ref{subhmf1} we plot  the subhalo  mass function  for GIF2
haloes at redshift  $z=0$, split according to the  final halo mass. We
considered all self-bound subhaloes  with at least $10$ particles, and
removed all subhaloes at halo-centric  distances $r < 0.05 R_v$, where
the subhaloes are difficult to  detect.  Note that the evolved subhalo
mass function is  not universal, but depends on the  mass of the final
host halo, with more massive  haloes hosting more subhaloes of a given
$m_{sb}/M_0$.

The fact that the evolved  subhalo mass function is not universal, but
rather  depends  on  host  halo   mass,  $M_0$,  was  first  shown  by
\citet{getal04}.   Using merger  trees constructed  with  the extended
Press-Schechter (EPS) formalism  \citep[e.g.,][]{lc93}, and adopting a
simple   model  for  the   average  mass   loss  rate   of  subhaloes,
\citet{vtg05}  were  able  to   reproduce  these  trends,  which  they
explained in  terms of (i)  the universality of the  unevolved subhalo
mass function,  and (ii)  the fact that  more massive  haloes assemble
later.   The  latter  implies   that  smaller  systems  accrete  their
satellites at  higher redshifts, when  the haloes are denser,  and the
dynamical times  are shorter.  This,  in turn, ensures  that dynamical
effects  that  promote mass  loss  are  more efficient.   Furthermore,
satellites that are  accreted earlier also are subjected  to mass loss
for a longer time. Both effects contribute to less massive host haloes
having less substructure at $z=0$.

Below we  will show  that the subhalo  populations in  our resimulated
haloes agree well with this  picture, and that indeed the average mass
loss rates are  higher at higher redshift. We will  also find that the
average mass  loss rates are  virtually independent of the  mass ratio
$m_{sb}(z)/M_v(z)$ between the subhalo and its host.

\section{Subhalo mass loss rates}
\label{massloss}

In this section  we estimate the subhalo mass  loss rate, modeling it
as a function  of (i) the instantaneous satellite  to host mass ratio:
$m_{sb}(z) / M_v(z)$, (ii) the mass of the host halo at redshift zero:
$M_0$, and (iii)  the cosmic time through the  redshift $z$.  For this
purpose  we  will  use   the  subhalo  population  identified  in  the
resimulations, as  haloes in this  sample have better force,  mass and
especially  time resolution  (88  snapshots between  redshift ten  and
zero) than  the cosmological GIF2  run.  Since mass loss  rates mostly
depend on the local environment  inside the host halo, the resimulated
sample will provide correct rates even if the haloes themselves do not
necessarily represent a fair sample for the given cosmological model.

\begin{figure}
\begin{center}
\includegraphics[width=\hsize]{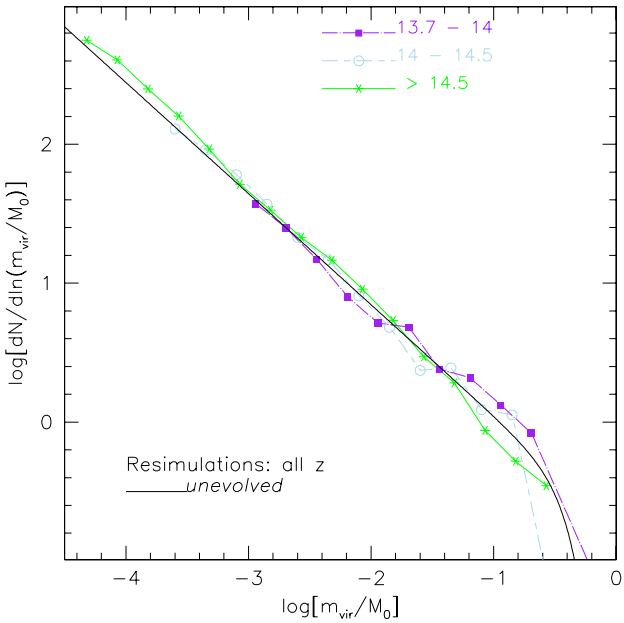}
\end{center}
\caption{\label{unshmfr03}Unevolved  subhalo  mass  function  for  the
  resimulated haloes.   We notice that the function  is independent on
  mass and well  described by the same function  fitting the GIF2 data
  (Figure~\ref{unshmf}). Haloes  are split  in three mass  bins. In
  the  figure  the  bounds  of  the  bin  are  expressed  in  unit  of
  $\log(M_{\odot}/h)$.}
\end{figure}

In Figure~\ref{unshmfr03} we show  the unevolved subhalo mass function
for  satellites  identified  in  the   merger  trees  of  the  set  of
resimulated  haloes;  host  haloes  are  split  in  three  mass  bins,
according  to  Table~\ref{tabsum}. As  for  the  GIF2 simulation,  the
unevolved  subhalo mass  function obtained  from the  resimulations is
well fit by eq.~\ref{unshmfeq}.

After  a satellite  enters  the  virial radius  of  the host,  various
dynamical effects, including  dynamical friction, tidal stripping, and
close encounters  with other subhaloes,  cause the subhaloes  to loose
mass,  and   may  eventually  result  in   their  complete  disruption
\citep[e.g.,][]{choi}.  The  (average) mass  loss rate of  dark matter
subhaloes is the direct link between the unevolved and evolved subhalo
mass   functions,   and  also   is   a   fundamental  ingredient   for
semi-analytical models  of galaxy  formation, as it  sets the  rate at
which satellite galaxies  merge with the central galaxy  in a halo, it
determines  the evolution  of  the mass-to-light  ratios of  satellite
galaxies, and  it regulates the  importance of stellar streams  in the
haloes of central galaxies.

\begin{figure*}
\begin{center}
  {\includegraphics[width=12cm]{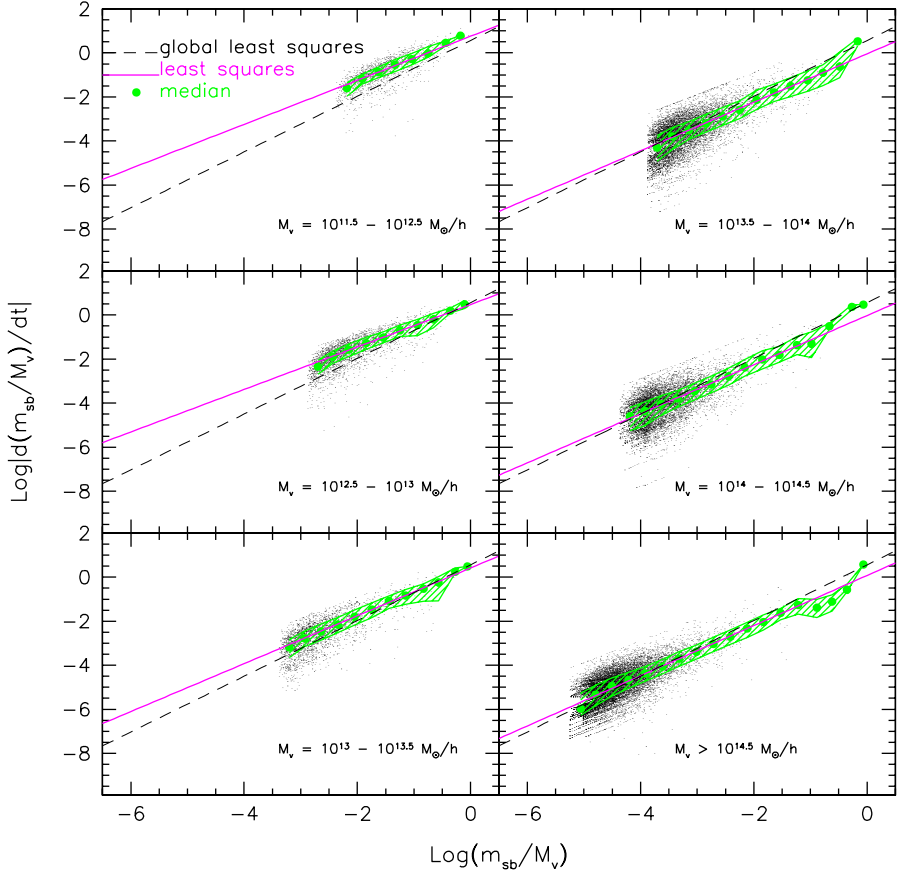}}  
  \caption{\label{masslossm} Subhalo mass loss rate. Each panel refers
  to a different bin in host halo virial mass at the redshift when the
  mass loss rate is computed.  The filled circles represent the median
  of points and the hatched region the quartiles. The thick solid line
  is  the  least  square  fit  to the  median  distribution  for  each
  panel. The thin  dashed line is the average least  squares for the different
  host halo masses.}
\end{center}
\end{figure*}

In  this section we  measure the  mass loss  rate experienced  by each
satellite.  In addition, using statistical averaging, we determine the
average mass loss  rate of satellites as a  function of the parameters
listed at  the beginning of this  section. We define  the average mass
loss  rate between  two successive  snapshots at  redshift,  $z_1$ and
$z_2$, as
\begin{equation}
  \frac{\mathrm{d}}{\mathrm{d}t}{\left(\frac{m_{sb}}{M_v}\right )}(z) =
  \frac{\dfrac{m_{sb}(z_{2})}{M_v(z_2)}- \dfrac{m_{sb}(z_{1})}
  {M_v(z_1)}}{t(z_{2})-t(z_{1})}\,,\,\,\,\,\,z_1 < z < z_2\,.
  \label{mlosseq}
\end{equation}
where the  values of $m_{sb}(z)$  and $M_v(z)$ are obtained  by linear
interpolation  of   the  values  at   $z_1$  and  $z_2$.    In  Figure
\ref{masslossm} we  plot the subhalo mass  loss rate as  a function of
the ratio $m_{sb}(z)/M_v$(z); Each panel refers to a different bin for
the mass $M_v(z)$ of the host halo.

The  green points  and  band in  each  panel indicate  the median  and
quartiles of  the distribution.  The  thick magenta line  represents a
least  squares fit  to the  median values  in each  panel; the  fit is
limited to the region where the median exhibits a linear behavior: we
excluded by  hand median values  for $m_{sb}/M_v$ close to  one, which
correspond to major  mergers and cannot be described  by a simple mass
loss model.  The  thin dashed black line, identical in  all panels, shows the
global least square fit obtained using the data from all panels.

The data  show a  clear linear relation  between $m_{sb}/M_v$  and its
time derivative, so we can write our model as:
\begin{equation}
  \log \left | \frac{\mathrm{d}(m_{sb}/M_v)}{\mathrm{d}t}\right | = a
  \log(m_{sb}/M_v) + b\,.
\label{mlossfit}
\end{equation}
Exponentiating this relation, and expanding the derivative on the LHS,
we obtain:
\begin{equation}
  \left | \frac{\dot{m}_{sb}}{M_v} - \frac{\dot{M}_v}{M_v}\frac{m_{sb}}{M_v} \right |=
  10^{b} \left (\frac{m_{sb}}{M_v} \right)^{a}\,.
\end{equation}
Due to  the large number of  snapshots in the  resimulations, the time
separation between two subsequent snapshots is always short: ${\rm d}t
\approx 0.1$ Gyr.  This is small  enough to assume a constant mass for
the host halo:  $\dot{M}_v = 0$. By doing so,  we obtain an expression
for the {\it specific} mass loss rate
\begin{equation}
  {\dot{m}_{sb} \over m_{sb}} = - \frac{1}{\tau} \left(
  \frac{m_{sb}}{M_v} \right)^{\zeta} \,,
\label{eqmassloss0}
\end{equation}
 where the free parameters $\tau(z,M_v) = 10^{-b}$ and $\zeta(z,M_v) =
a  -  1$  might  in  principle  depend  both  on  cosmic  time  (or,
equivalently, redshift  $z$) and  on the virial  mass $M_v(z)$  of the
host halo at that time.   The negative sign arises from the mass loss
of  the  satellites.   Note  that  this specific  mass  loss  rate  is
identical to that used by \citet{vtg05}.
\begin{figure}
\centering
\includegraphics[width=8cm]{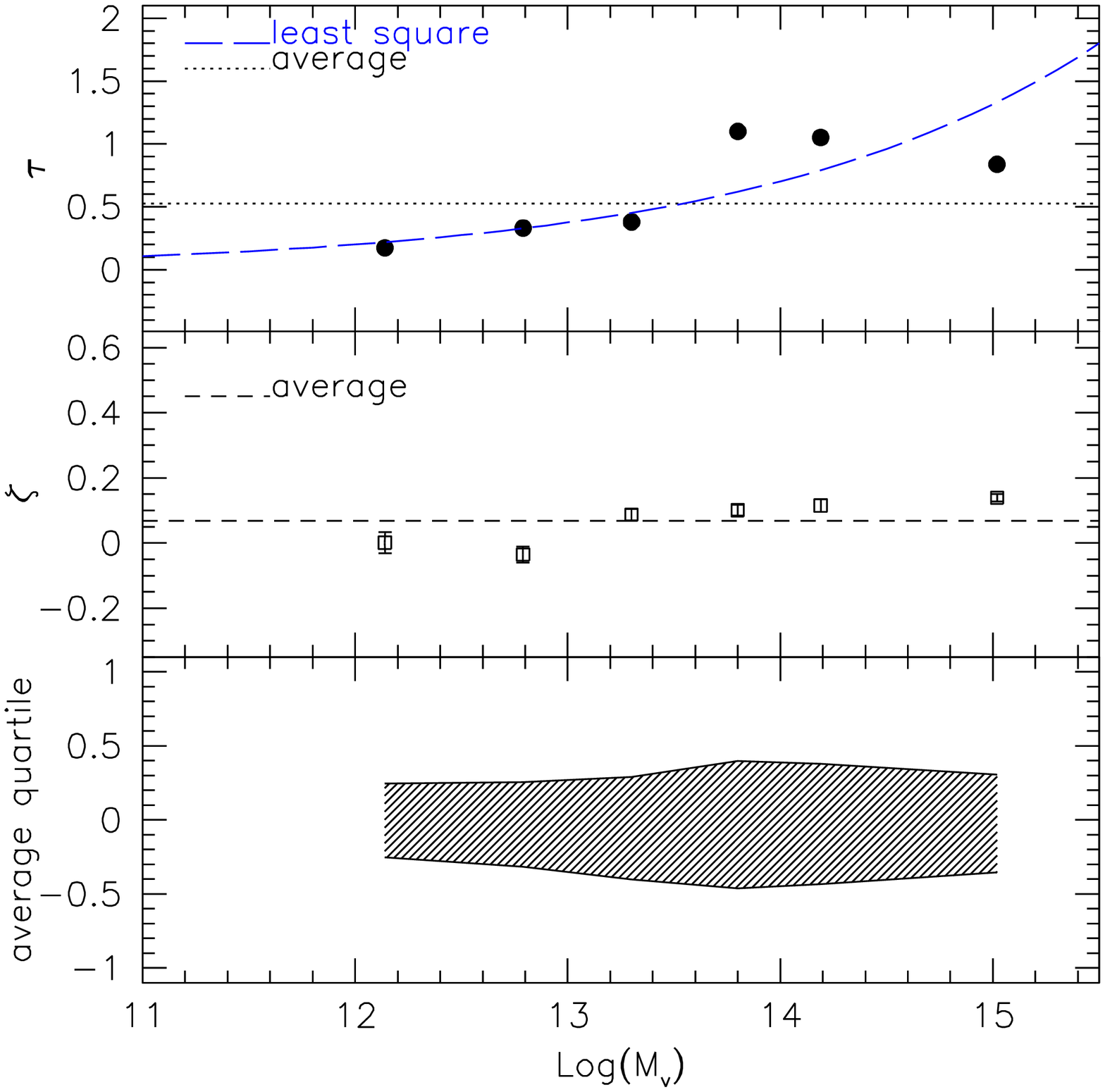}
\caption{\label{szmass}Dependence of the  fit parameters of the Figure
  \ref{masslossm} on  the host halo  virial mass. The top  panel shows
  the time scale  of the mass loss rate $\tau=10^{-b}$. The  average and the
  least square  fit of the data points have been computed  on the plane  ($b$, $M_v$). In
  the central panel we show the dependence of parameter $\zeta=a-1$ on
  $M_v$.  In  the bottom  panel we  show the spread  of the  first and
  third quartiles around  the median, averaged over the  six panels of
  Figure   \ref{masslossm}  (see   the  main   text  for   a  detailed
  explanation).}
\end{figure}

Figure~\ref{szmass} shows  how the time scale of the mass loss rate, $\tau=10^{-b}$, and  
$\zeta=a-1$, as measured from the data  shown in Figure~\ref{masslossm}, depend on the
virial  mass,  $M_v$, of  the  instantaneous  host  halo.  Error  bars
reflect the  usual uncertainty on  the coefficients obtained  from the
least square fitting. The slope is found to be independent of the mass
of the host halo, with a best fit value of $a=1.07 \pm 0.03$ ($\zeta =
0.07  \pm 0.03$). This  implies that  the specific  mass loss  rate is
almost independent  of the instantaneous mass  ratio $m_{sb}/M_v$.  On
the other  hand, the zero point, $b$,  is found to be  larger for less
massive haloes.

In  order to  show  the typical  spread  of points  in  each panel  of
Fig.~\ref{masslossm}  around  each  median,  in the  bottom  panel  of
Figure~\ref{szmass} we show the average (over the six panels of Figure
\ref{masslossm})  of the  differences  between each  quartile and  the
median itself. We see that on  average fifty percent of the points lay
roughly within a distance $\log y = \pm 0.3$ from the median; that is,
typical mass  losses deviate  from their median  value by less  than a
factor of two.
\begin{figure*}  
\begin{center}
  {\includegraphics[width=12cm]{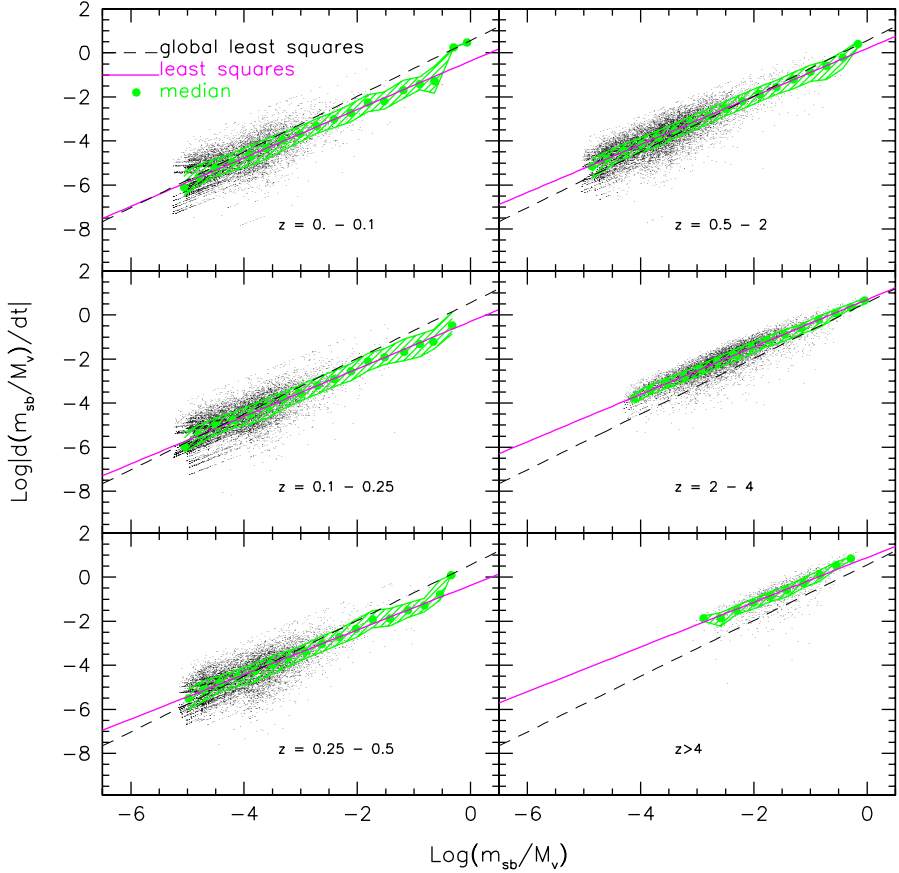}}  
  \caption{\label{masslossz}Subhalo mass loss rate. Each panel refers
    to a different bin in the redshift at which the mass loss rate is
    computed. Symbols and lines are as in Figure~\ref{masslossm}.}
\end{center}
\end{figure*}

\begin{figure}
\centering
\includegraphics[width=8cm]{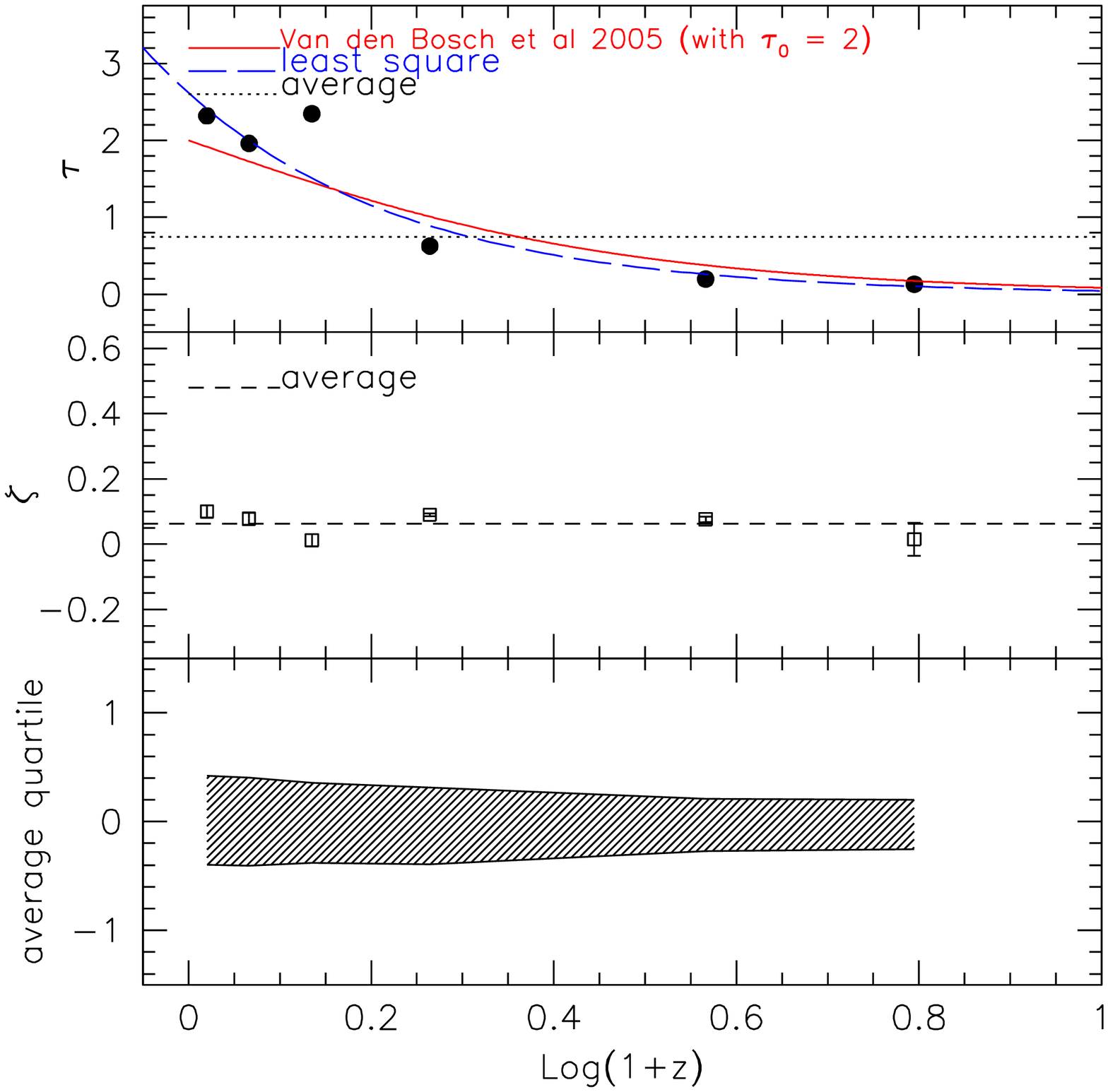}
\caption{\label{szred} Time scale of the mass loss rate and $\zeta$ in
   term of  the redshift at  which the subhaloes are  experiencing mass
   loss (Figure  \ref{masslossz}. The average and the  least squares fit of
   the top  panel were  computed on the  plane ($b$, $z$).  The bottom
   panel shows  the average  first and third  quartile for  the median
   distribution   in  each   panel   of  the   Figure~\ref{masslossz},
   constructed as previously described in the main text.}
\end{figure}

In Figure~\ref{masslossz}  we plot the  subhalo mass loss  rate versus
the ratio $m_{sb}(z)/M_v(z)$, now  binned according to the redshift at
which the mass loss rate  is calculated.  Medians, quartiles and lines
are as in Figure~\ref{masslossm}.  The time scale $\tau$ and $\zeta$ for the six
panels  are  shown  in  Figure~\ref{szred}, plotted  versus  the  mean
redshift of  each of  the six  bins; in the  bottom panel  the average
quartile distribution for each fit (as explained above) is shown.  The
red  solid  curve superimposed  to  the trend  in  zero  point is  the
equation
\begin{equation}
  \tau(z) = \tau_{0} \left[
  \frac{\Delta_{v}(z)}{\Delta_{0}}\right]^{-1/2} \left[
  \frac{H(z)}{H_0}\right]^{-1}\,,
  \label{eq:tauz}
\end{equation}
with $H(z)$ the Hubble  constant at redshift  $z$, and with  $\tau_0 =
2.0\,\mathrm{Gyr}$.

This equation was proposed by \citet{vtg05} and describes the redshift
dependence  of mass  loss  rates obtained  under  the assumption  that
$\tau$  is proportional  to the  dynamical time  $t_{\rm  dyn} \propto
\rho_v^{-1/2}(z)$,  taking   into  account  that,   according  to  the
spherical  collapse  model,  the  average density  within  the  virial
radius, $\rho_v$ is independent of  halo mass at fixed redshift.  This
means  that we  can write  $\tau(M_v,z) =  \tau(z)$. The  red  line in
Figure~\ref{szred} shows that indeed  this provides a good description
of the measured mass loss rates.

%
%

\begin{figure}  
\begin{center}
  \includegraphics[width=\hsize]{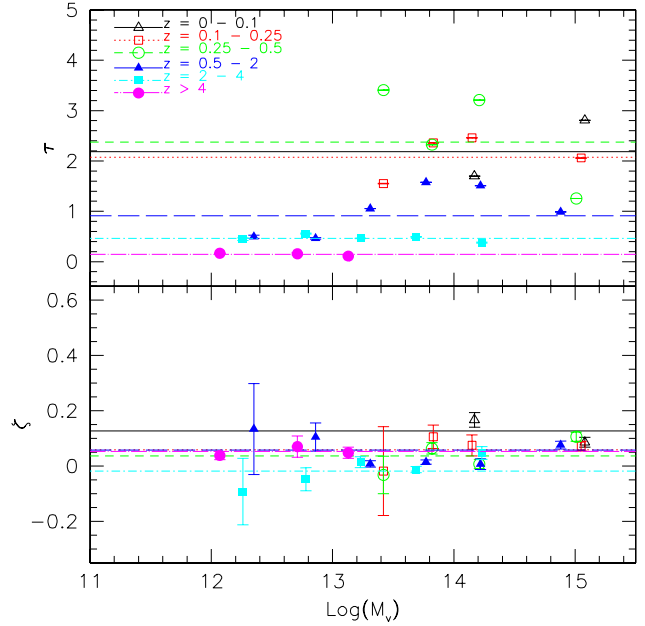}
  \caption{\label{vaimdz}Time scale of the  mass loss rate and $\zeta$
  versus  host mass,  for six  fixed redshift  bins --  represented by
  different  data points.   The  horizontal lines,  with various  line
  type, show the average $b=-\log(\tau)$ and $\zeta$ for each redshift bin.}
\end{center}
\end{figure}

Note, though, that Figure~\ref{szmass}  suggests that the average mass
loss rates also depend on host  halo mass.  In order to reconcile this
with the  claim that  the zero-point is  independent of  $M_v$, recall
that, on average, more massive haloes assemble (and thus accrete their
satellites)  earlier   than  less  massive   haloes.   Therefore,  the
different panels of Figure~\ref{masslossm} actually refer to different
average redshifts, with larger  $M_v$ corresponding to a lower average
redshift.  Consequently, the `apparent' mass dependence evident in the
upper  panel of  Figure~\ref{szmass}  is merely  a  reflection of  the
redshift  dependence   described  by  equation   (\ref{eq:tauz}).   To
demonstrate  this  we now  split  the data  points  of  each panel  of
Figure~\ref{masslossz} in different subsets,  according to the mass of
the host halo.  Figure~\ref{vaimdz}  shows the average slopes and zero
points obtained  for these  subsets using least-squares  fitting. This
clearly shows that the characteristic  time scale for mass loss (given
by the  zero point) is independent of  the host mass  $M_v(z)$ at fixed
redshift,  in accord  with  equation (\ref{eq:tauz}).   

Thus, to good approximation, the average mass loss rate of dark matter
subhaloes depends  only on the density  of the host halo,  and thus on
redshift  (or cosmic time),  but not  on the  mass of  the host  halo. 
Furthermore, since the best-fit value  of $\zeta$ is close to zero, to
good  approximation subhalo  masses  decay exponentially\footnote{this
  follows  from a simple  integration of  equation (\ref{eqmassloss0})
  with $\zeta=0$} according to
\begin{equation}
m_{sb}(t) = m_{v} \exp\left[-{t - t_m \over \tau(z)}\right]\,,
\end{equation}
where $m_v$  is the mass  of the satellite  at the time  of accretion,
$t_{m}$,  and  $\tau(z)$  is  given by  equation~(\ref{eq:tauz})  with
$\tau_0 = 2.0\,\mathrm{Gyr}$.

\section{Monte Carlo simulation}
\label{mcsim}

In this section we compare  our results to those of \citet{vtg05}, and
we use their  Monte Carlo method to check  the self-consistency of the
results  presented  above,  i.e.,  we check  whether  the  (universal)
unevolved subhalo mass function, combined with the satellite accretion
times  and the  average mass  loss  rates, can  reproduce the  evolved
subhalo mass functions presented in Section~\ref{subhaloes}.

The Monte-Carlo  method  of \citet{vtg05}  starts  by constructing EPS
merger  trees  using the  method  described  in \citet{w02} \citep[see
also][]{sk99}.   These   merger trees are  then   used to register the
accretion  times  and  masses  of  satellites   merging onto the  main
progenitor.  Starting from  these inputs, \citet{vtg05} then proceeded
as follows. In  between two time-steps, they evolve  the masses of the
subhaloes using equations~\ref{eqmassloss0} and~\ref{eq:tauz}. The two
free parameters, $\tau_0$ and  $\zeta$   were tuned to reproduce   the
subhalo mass function of  massive, cluster sized haloes  obtained from
numerical simulations   by  \citet{getal04},    \citet{delucia04}  and
\citet{tmy04}.  This resulted in   $\tau_0  = 0.13$ Gyr  and  $\zeta =
0.36$,   which differs substantially  from  the results obtained here:
$\tau_0=2.0$ Gyr and $\zeta  =0.06$.  The reason for this  discrepancy
owes to  the  use of  EPS merger  trees, as  opposed to  merger  trees
extracted from numerical simulations.  In fact, the  unevolved subhalo
mass  function obtained  by  \citet{vtg05}  differs significantly from
that shown in Figures~\ref{unshmf}  and~\ref{unshmfr03}, in that it is
significantly  higher,  and with  a different slope    at the low mass
end\footnote{It is noteworthy, though, that the EPS formalism predicts
  that  the   unevolved subhalo   mass function   is universal,   i.e.,
  independent of the host mass,  in good agreement with the simulation
  results presented here.}.   Consequently, in order to  reproduce the
subhalo  mass functions      obtained  from  numerical    simulations,
\citet{vtg05} had to adopt higher mass loss rates (i.e., a lower value
for $\tau_0$, and a different mass dependence (i.e., a different value
for $\zeta$).

\begin{figure}  
\begin{center}
  \includegraphics[width=\hsize]{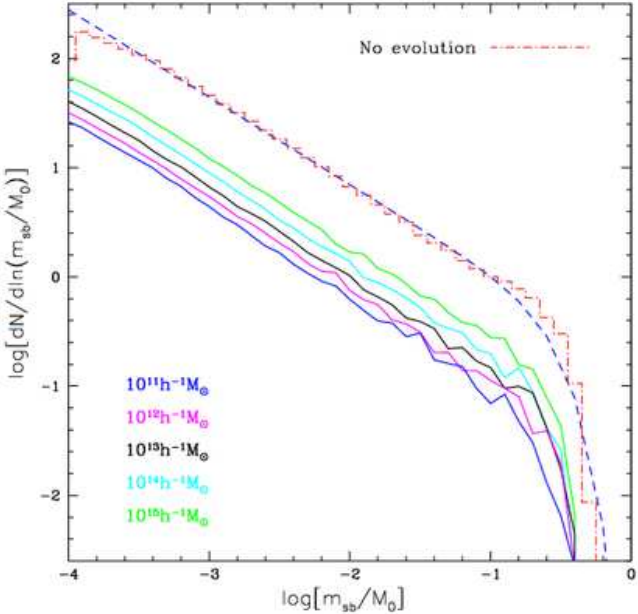}
  \caption{The dotted histogram show the mass accreted by the main
    branch in the Monte Carlo merger tree with the overplotted equation
    (\ref{unshmfeq}). The solid lines represent the subhaloes mass
    function obtained evolving the mass accreted by the main
    progenitors of different present day $M_0$-halo.\label{vdb1}}
\end{center}
\end{figure}

The  fact that  EPS merger  trees  predict an  unevolved subhalo  mass
function that  differs significantly  from that obtained  in numerical
simulations, should  not come entirely  as a surprise. After  all, the
construction  of EPS  merger trees  relies on  the  spherical collapse
model \citep[see][]{lc93,sk99}.  However,  in reality, the collapse of
dark matter haloes is influenced  by the surrounding tidal force field,
making    the   collapse    ellipsoidal,    rather   than    spherical
\citep[e.g.,][]{st99,smt01,st02}.   As   shown  by  \citet{st02},  the
conditional  and  unconditional  mass  functions are  different  under
ellipsoidal   collapse  conditions   than  under   spherical  collapse
conditions, which  has important consequences for the  accuracy of the
EPS merger trees. For instance,  the halo formation times predicted by
EPS  are  systematically offset  from  those  obtained from  numerical
simulations    \citep{lc93,som00,w02,wetal02,giocoli07},   while   the
average  mass  of  the  main  progenitor  is  typically  overestimated
\citep{som00}.

To perform  the self-consistency  check mentioned above,  we therefore
use  the same  Monte-Carlo method  as \citet{vtg05},  but  we randomly
remove satellite-branches from the merger tree with a probability
\begin{equation}
P_{\rm reject} = {n_{\rm sim}(m_v/M_0) \over n_{\rm EPS}(m_v/M_0)}
\end{equation}
where $n_{\rm sim}$  and $n_{\rm EPS}$ are the  unevolved subhalo mass
functions obtained from the simulations and from the EPS merger trees,
respectively.   This  ensures that  the  Monte  Carlo  method uses  an
effective, unevolved  subhalo mass function that is  identical to that
of equation~(\ref{unshmfeq}).

As    in \citet{vtg05} we  evolve the   masses  of the subhaloes using
equations~\ref{eqmassloss0}   and~\ref{eq:tauz}  with $\tau_0=2.0$ Gyr
and   $\zeta =0.06$, which   are   the best-fit   values obtained   in
section~\ref{massloss}.  The resulting evolved subhalo mass functions,
for five different masses  of the present-day host  halo, are shown in
Figure~\ref{vdb1}, together  with the unevolved subhalo  mass function
obtained using   the rejection scheme  outlined  above (and   which is
independent of the host halo mass). Each evolved subhalo mass function
is  the   average    obtained from   2000   merger   tree realizations
\citep[see][for details]{vtg05}. A comparison with the evolved subhalo
mass functions  obtained from our numerical  simulations, and shown in
Figure~\ref{vdb1}, shows good  agreement.   This  indicates that   the
evolved subhalo   mass   functions   are   self-consistent  with   the
(universal) unevolved subhalo  mass function and  the  simple form for
the average mass loss rate obtained in this paper.

\section{Summary and Conclusions}
\label{conclusion}

In  this paper  we have  studied  the mass  loss rate  of dark  matter
subhaloes  using  a set  of  high  resolution  $N$-body simulation  of
structure formation.  Haloes were  followed backward in time along the
main  branch of  their merging  history  tree.  At  each snapshot  the
satellites  accreted by the  main branch  were identified.   We showed
that the mass function  of accreted satellites (unevolved subhalo mass
function) is universal, that is, it does not depend on the present day
host halo  mass $M_0$,  and we presented  a fitting function  for this
distribution.

We then followed each accreted  satellite forward in time, snapshot by
snapshot, computing   its self-bound mass and its   mass loss rate. We
found that    the  expression for the   mass   loss rate   proposed by
\citet{vtg05} is consistent  with $N$-body simulations, and  excellent
agreement is obtained  if the value  with $\tau_0 = 2.0\,\mathrm{Gyr}$
is  taken.  In addition,  we find that the average  mass  loss rate is
virtually   independent of the  instantaneous  mass ratio $m_{sb}/M_v$
between the subhalo  and  its host halo.  This  differs  substantially
from the best-fit mass  loss rate parameters obtained by \citet{vtg05}
using EPS merger trees.  In particular, \citet{vtg05} obtained
$\tau_0=0.13\,\mathrm{Gyr}$,   and   a   significant     dependence on
$m_{sb}/M_v$. The  reason for this discrepancy  is that  the unevolved
subhalo mass  function  of EPS merger trees   is too high,  so that  a
higher  mass loss rate was inferred  to be consistent with the evolved
subhalo mass functions in numerical simulations.

With  an unevolved  subhalo mass  function that  is universal,  and an
average mass loss rate  that is virtually independent of $m_{sb}/M_v$,
it becomes straightforward to  understand why less massive haloes have
evolved subhalo mass functions with a lower normalization. This simply
owes  to the  fact that  less massive  haloes assemble  earlier, which
implies that  they accrete their satellites earlier.  At earlier times
the  mass loss rate  is higher,  because the  dynamical times  of dark
matter haloes  are shorter.  In addition, a  subhalo that  is accreted
earlier  is subjected to  mass loss  for a  longer period.  Both these
effects  contribute to  the fact  that less  massive haloes  have less
substructure.
 
The  present description does  not consider  the possible  presence of
subhaloes within  subhaloes, accreted along  the tree of  each present
day  subhalo.   In  a  follow-up   paper  (Giocoli  et  al.  2008,  in
preparation) we  will investigate this issue in  detail, comparing the
populations of subhaloes found using different techniques.

\section{Acknowledgments}
This work has been partially supported by Italian PRIN and ASI.

Thanks to Ravi K. Sheth, Robert Smith and Gao Liang for useful discussions 
and comments. Thanks also to Marco Comparato and Ugo Becciani for useful assistance with
VisIVO  \citep{comparato}:   a  visualization   interface  for
numerical simulation.

\bibliographystyle{mn2e}

\label{lastpage}
\end{document}